\documentclass[iop,revtex4]{emulateapj}

\newcommand{\degree}{^\circ}

\usepackage{amsmath}
\usepackage{rotating}

\shorttitle{AU Mic Disk Color}
\shortauthors{Lomax et al.}

\begin{document}


\title{Optical Coronagraphic Spectroscopy of AU Mic: Evidence of Time Variable Colors?}


\author{Jamie R. Lomax}
\affil{Department of Astronomy, Box 351580, University of Washington, Seattle, WA 98195, USA}
\email{jrlomax@uw.edu} 

\author{John P. Wisniewski}
\affil{Homer L. Dodge Physics and Astronomy Department, University of Oklahoma, 440 W Brooks St., Norman, OK 73019, USA}
\email{wisniewski@ou.edu}

\author{Aki Roberge}
\affil{Exoplanets \& Stellar Astrophysics Laboratory, NASA Goddard Space Flight Center, Code 667, Greenbelt, MD, 20771, USA}

\author{Jessica K. Donaldson}
\affil{Department of Terrestrial Magnetism, Carnegie Institution of Washington, Washington DC 20015, USA}

\author{John H. Debes}
\affil{Space Telescope Science Institute, 3700 San Martin Dr., Baltimore, MD 21218, USA}

\author{Eliot M. Malumuth}
\affil{Adnet Systems, Inc., 6720B Rockledge Dr., Suite 504, Bethesda, MD, 20817 USA}

\author{Alycia J. Weinberger}
\affil{Department of Terrestrial Magnetism, Carnegie Institution of Washington, Washington DC 20015, USA}



\begin{abstract}
We present coronagraphic long slit spectra of AU Mic's debris disk taken with the STIS instrument aboard the \textit{Hubble Space Telescope} (HST). Our spectra are the first spatially resolved, scattered light spectra of the system's disk, which we detect at projected distances between approximately 10 and 45 AU. Our spectra cover a wavelength range between 5200 and 10200 \AA. We find that the color of AU Mic's debris disk is bluest at small (12-35 AU) projected separations. These results both confirm and quantify the findings qualitatively noted by \cite{Krist}, and are different than IR observations that suggested a uniform blue or gray color as a function of projected separation in this region of the disk. Unlike previous literature that reported the color of AU Mic's disk became increasingly more blue as a function of projected separation beyond $\sim$30 AU, we find the disk's optical color between 35-45 AU to be uniformly blue on the southeast side of the disk and decreasingly blue on the northwest side. We note that this apparent change in disk color at larger projected separations coincides with several fast, outward moving ``features'' that are passing through this region of the southeast side of the disk. We speculate that these phenomenon might be related, and that the fast moving features could be changing the localized distribution of sub-micron sized grains as they pass by, thereby reducing the blue color of the disk in the process. We encourage follow-up optical spectroscopic observations of the AU Mic to both confirm this result, and search for further modifications of the disk color caused by additional fast moving features propagating through the disk. 
\end{abstract}


\section{Introduction}

AU Mic is a 23 $\pm$3 Myr old \citep{bin14,mal14,mam14}, nearby (9.9 pc), M1Ve star that is a member of the $\beta$ Pic moving group \citep{Barrado,Perryman,Zuckerman}. The star is surrounded by a debris disk that extends out to 210 AU, which appears close to edge-on from our point of view \citep{Kalas,Liu,Krist}. While several early searches for planets in the system yielded null results \citep{Hebb,Metchev}, \citet{pla17} recently presented radial velocity data that suggested the presence of two Jovian-mass planets at semi-major axis separations of $\sim$0.3 and 3.5 AU.

AU Mic exhibits numerous signatures of activity common for its youth. For example, \cite{Aki2} observed flares from AU Mic in both their \textit{FUSE} and \textit{HST/STIS} data. \cite{Hebb} monitored AU Mic for 28 continuous nights in 2005 with the CTIO 1-m telescope in the optical and found that the system exhibited a 4.847 day period, which they attributed to star spots, and underwent three flares, while \cite{UVFlares} observed several flaring events from the system with \textit{HST/STIS} that lasted between 10 s and 3 min. This is common for flares on M-dwarfs, which last on the order seconds to many hours and result in a significant increase in the stars' flux at NUV and blue optical wavelengths (see e.g. \citealt{kow13}). \cite{Mitra} found that the flares in the UV typically precede those in the X-ray wavelengths using the \textit{XMM-Newton} EPIC and OM cameras. Chandra observations also detected a small flare from the system \citep{Linsky}, and simultaneous observations with \textit{XMM-Newton} and the VLA showed that the largest X-ray flares also have a radio counterpart \citep{XrayRadioFlares}. A high energy (10$^{36}$ ergs) flare was also observed in EUVE observations of the system, and interpreted to be accompanied by a coronal mass ejection \citep{cul94}.

AU Mic's debris disk has been spatially resolved across a wide range of wavelength regimes, including optical to near-IR scattered light \citep{Kalas,Liu,Krist,Fitzgerald,Metchev,sch14,wang,diskvar}, optical polarized scattered light \citep{Graham}, far-IR and sub-mm \citep{mat15}, and mm \citep{MacGregor} wavelengths, helping to trace the spatial distribution of a variety of grain size populations. The system appears to have a `birth ring' of material near 43 AU that forms micron sized particles from collisions between larger bodies (see e.g. \citealt{Strubbe, Graham,MacGregor,mat15}), as well as an extended halo comprised of approximately micron-sized grains \citep{Kalas,mat15}.  Recent analysis of ALMA data suggest that there are millimeter sized grains at separations of $\leq 3$ AU from the central star and an outer dust belt near the birth ring, the combination of which might be architecturally similar to the asteroid belt and Kuiper Belt in our own solar system \citep{MacGregor}. Although, recent modeling of the ALMA data suggests that coronal heating may be responsible for the point source excess at millimeter wavelengths instead of an inner dust belt \citep{Cranmer}. The inner ($<$ 30 AU) disk appears to be largely devoid of small, micron-sized grains \citep{Strubbe}, while analysis of the polarimetric properties of the disk led \citet{Graham} to suggest that its grains were highly porous.  

There is a rich history of attempting to diagnose the optical and near-IR colors of AU Mic's debris disk. When comparing \textit{HST} F606W and F435W coronagraphic imagery to each other, \cite{Krist} found the disk's color to be gray. However, a comparison of the F435W and F814W data show that AU Mic's disk is blue relative to the central star, particularly at large radii (30 to 60 AU). \cite{Krist} also qualitatively suggested that the disk looks increasingly blue inwards of 30 AU, but cautioned that PSF residuals may have biased that result. The increasingly blue color of the outer disk persists in the near-IR and when comparing \cite{Krist}'s optical \textit{HST} data to near-IR datasets \citep{Fitzgerald,Metchev}. Nevertheless, the increasingly blue trend at small projected separations has not been reproduced \citep{Fitzgerald}. It's thought that the radiation pressure from the central M star, which is lower than that of more massive stars, is ineffective at clearing out grains, allowing small grains to persist longer than typically expected for a debris disk \citep{Krist}. Instead, it is believed that the evolution of the grain populations within AU Mic's disk is more strongly affected by a combination of their porosity and the central star's wind, which may replace radiation pressure as the dominate mechanism for clearing out small grains. \citep{Augereau,mat15}. 

Local density enhancements and deficits have been discovered within scattered light images of the disk \citep{Liu,Krist,Fitzgerald,sch14} that are consistent with ongoing or recent planet formation in the system.  Moreover, recent analysis of multi-epoch imagery of the disk has revealed clear evidence of large scale features at projected separations of 10-60 AU on the southeast side of the disk that change location as a function of time \citep{diskvar}. These features are moving between 4 and 10 km s$^{ -1}$ and appear to be either in highly elliptical orbits, or unbound. While the fundamental mechanism responsible for producing this new observational phenomenon remains unclear, \citet{diskvar} did speculate that highly energetic episodic flares may be interacting with the disk material. Given the newly discovered dynamic nature of AU Mic's disk, one might expect that the disk's color could exhibit time variability if the fundamental mechanism driving the fast moving features also changed the local grain size distribution. In fact, it has been theoretically shown that coronal mass ejections, one of the possible causes of the fast moving features in AU Mic's disk, could remove infrared emitting dust from disks around active stars on timescales of days \citep{Osten}.

In this paper, we investigate the color of AU Mic's debris disk using the \textit{HST/STIS} coronagraphic spectroscopy mode, which makes use of a long slit and occulting bar to produce optical spectra of the disk as a function of radius from the central star. Despite coronagraphic spectroscopy's relative power to measure the location dependent color of a disk at a higher spectral resolution than simple multi-band photometry, there is only one previously published dataset in which this technique has been used on a disk system (TW Hya; \citealt{Aki}). Therefore, in Sections 2 and 3 we describe our observing strategy and data reduction procedures in detail (see also \citealt{Aki} for a comprehensive description of this technique). In Section 4 we present our spatially resolved disk color spectra. In Section 5 we describe the surface brightness and disk color profiles extracted from our dataset and compare them to datasets already in the literature. We discuss the implications of our findings in Section 6 and summarize our findings in Section 7.

\section{Observations}

The \textit{Hubble Space Telescope} (\textit{HST}) observed AU Mic and a point spread function (PSF) reference star, GJ 784, on 2012 July 29 as part of program GO-12512. All observations were taken using the G750L grating with the \textit{STIS} instrument and cover the 5240-10270 \AA\space wavelength range.

Our observing strategy was similar to that of \cite{Aki} and produces spatially resolved coronagraphic spectroscopy of AU Mic's disk by making use of the 0\farcs 86 fiducial occulting bar attached to the STIS long slits. Our AU Mic observations consist of a sequence of exposures that start with peak ups using a narrow slit (0.2 $\times$ 0.05ND) in both the wavelength (x) and spatial (y) directions of the detector to accurately position the star after the initial acquisition. These are followed by an unocculted spectrum of AU Mic, and a lamp flat using the 52"$\times$0\farcs 2 slit.  Next, a series of exposures using the 52"$\times$0\farcs 2F2 slit are taken (hereafter ``fiducial exposures", ``fiducial data", or ``fiducial spectra"). This is the same slit as the one used for the unocculted spectrum, but it places the 0\farcs 86 wide fiducial bar over the central star. We repeated this sequence of exposures for the PSF star GJ 784.  During our first three \textit{HST} orbits, we collected an unocculted point source spectrum and three fiducial spectra of AU Mic, as well as a lamp flat. The lamp flat, the unocculted AU Mic spectrum, and one fiducial spectrum were collected during the first orbit, and we obtained two additional fiducial spectra during the following two orbits. Therefore, the exposure time of our first fiducial spectrum of AU Mic (OBPZ05030) is significantly shorter (1520 s) than the following two spectra (OBPZ05040 and OBPZ05050 respectively; 2780 s). During our fourth and final orbit we obtained an unocculted point source spectrum and one fiducial spectrum of the PSF star, as well as an additional lamp flat. Table \ref{Obs} lists the start times, exposure times, and apertures used for our observations.

\begin{center}
\begin{figure}
\centering
\includegraphics[width=0.5\textwidth]{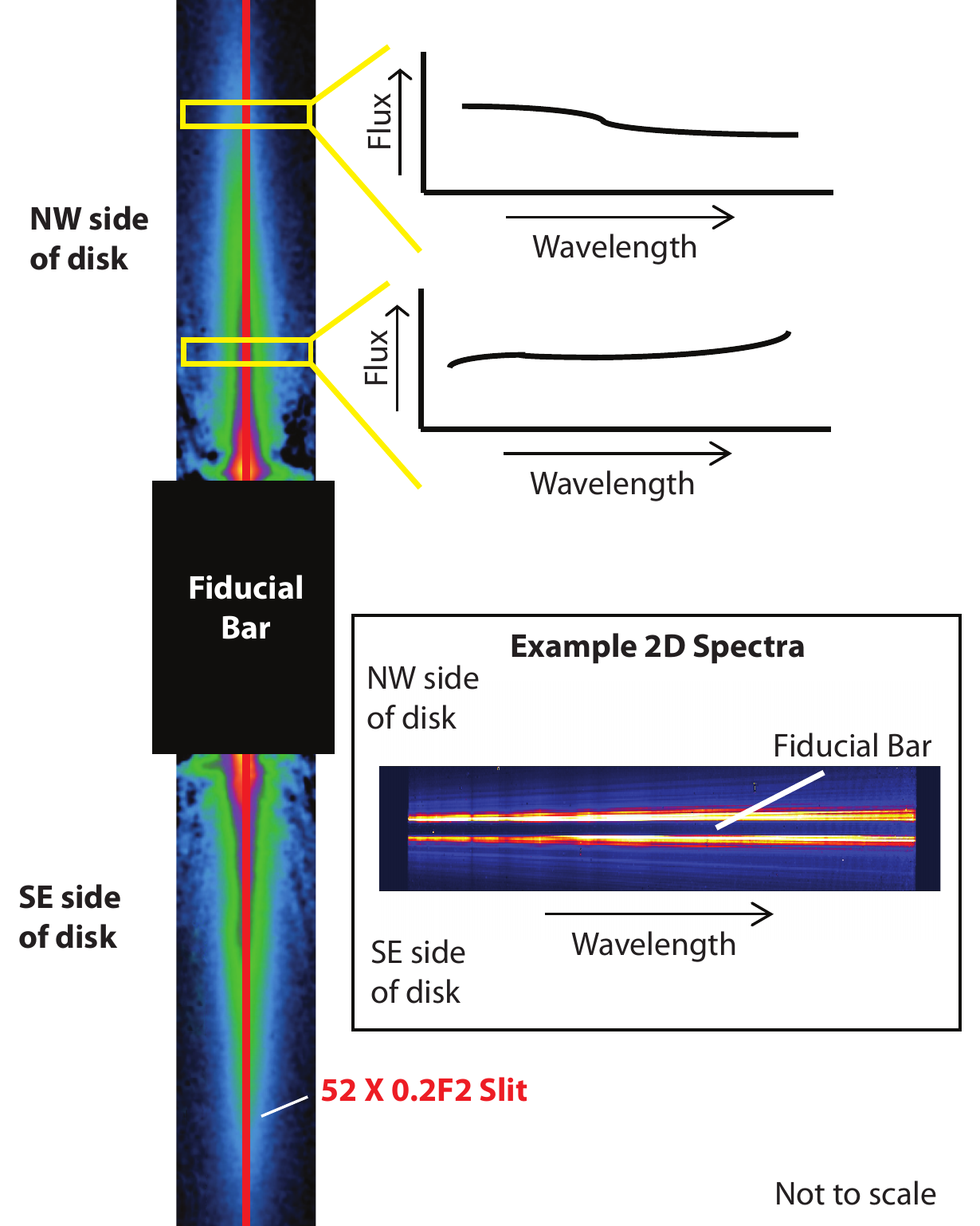}
\caption{The \textit{HST} instrumental setup for coronagraphic spectroscopy. The 52"$\times$0\farcs 2F2 slit (red line) is aligned so that it is parallel to AU Mic's debris disk (image credit: Fitzgerald et al. 2007). The slit's fiducial bar (black box) acts as a coronagraph and blocks out light from the system's central star. Each slice of the resulting data (perpendicular to the slit, yellow box) is a spectrum of the disk. The boxed inset shows an example observation taken with this observing method. The location of each side of the disk, as well as the fiducial bar are marked. }
\label{InstConfig}
\end{figure}
\end{center}

In the case of AU Mic, the position angle of the slit, and therefore the roll angle of \textit{HST}, was chosen so that it was aligned with the disk. For all of our AU Mic fiducial observations the slit was oriented at 307.1$\degree$, which places it at very close to the same angle as the disk (Kalas et al. 2004 report the position angle of the disk as 124$\degree\pm$2$\degree$ for the south east side and 310$\degree\pm$1$\degree$ for the north west side). This causes the signal from the disk to be dispersed along the x-axis of the exposure so that each row of pixels represents a spectrum from a different spatial location of the disk (i.e. the x-axis is the wavelength direction and the y-axis is the spatial direction), as shown in Figure \ref{InstConfig}. Therefore, data on the +y (-y) side of the fiducial bar are from the north west (south east) side of the disk (hereafter NW and SE). In the spatial direction the pixel size is 0\farcs 051. We note that no matter the roll angle of \textit{HST}, the slit is always oriented away from the diffraction spikes \citep{STIS}.

\begin{center}
\begin{deluxetable*}{lccccc}
\tablecolumns{6}
\tablewidth{0pt}
\tablecaption{Observation Log \label{Obs}}
\tablehead{
\colhead{Obs ID} &\colhead{Target} &\colhead{Start Time (UT)} &\colhead{Exposure Time (s)} &\colhead{Aperture} &\colhead{Comments} }
\startdata
OBPZ05010 &	AU Mic &12:15:09 &0.6 &52 $\times$ 0.2 &Point source spectrum\\
OBPZ05020 & Lamp Flat &12:17:21 &25 &52 $\times$ 0.2 &\\
OBPZ05030 & AU Mic &12:20:21 &1520 &52 $\times$ 0.2F2 &Fiducial spectrum\\
OBPZ05040 & AU Mic &13:32:30 &2780 &52 $\times$ 0.2F2 &Fiducial spectrum\\
OBPZ05050 & AU Mic &15:08:20 &2780 &52 $\times$ 0.2F2 &Fiducial spectrum\\
OBPZ06010 & GJ 784 &17:30:27 &0.4 &52 $\times$ 0.2 &Point source spectrum, PSF star\\
OBPZ06020 & Lamp Flat &17:32:39 &25 &52 $\times$ 0.2 &\\
OBPZ06030 & GJ 784 &18:20:15 &1678 &52 $\times$ 0.2F2 &Fiducial spectrum, PSF star
\enddata
\tablecomments{All data were taken with the STIS CCD using the G750L grating.}
\end{deluxetable*}
\end{center}

\section{Data Reduction} 

As described in Section 2, the AU Mic and PSF star exposures were obtained with the 52"$\times$0\farcs 2F2 slit, which is an unsupported STIS observing mode. Therefore, we performed our own data reduction and calibration on both the point source and fiducial datasets following the prescription laid out by \cite{Aki}, who used the same data collection technique on the TW Hya protoplanetary disk system. The general data reduction steps, which we discuss in detail below, include defringing, wavelength calibration, and flux calibration for both the point source and fiducial data. We also perform hot and cold pixel correction and PSF subtraction on our fiducial datasets. We explored processing a combined version of the AU Mic fiducial data, which we obtained by adding the raw exposures from all the orbits together, but we found that this reduced our ability to successfully defringe the data. Therefore, we processed the individual orbits of the AU Mic fiducial data separately. 

We began our data reduction by running the raw point source and fiducial exposures through the \texttt{STSDAS IRAF} \textit{prepspec} task, which subtracts bias and dark frames, divides by the pixel to pixel flat, subtracts overscan levels, and rejects cosmic rays.

\subsection{Defringing, Wavelength Calibration, and Flux Calibration}

Interference between reflections off the front and back surfaces of the STIS CCD causes fringing in all (point source and fiducial) our spectra at long wavelengths (see also Roberge et al. 2005). We defringed all of our data by using our contemporaneous lamp flats and the \texttt{STSDAS} \textit{normspflat}, \textit{mkfringeflat}, and \textit{defringe} \texttt{IRAF} tasks. This allows us to create and apply a flat field with fringes that have been shifted and scaled to match those in the point source and fiducial spectra. However, some nominal amount of fringing still remains, particularly above the fiducial bar on the NW side of the disk, and at wavelengths above 9,500 \AA.

\begin{center}
\begin{figure*}
\centering
\includegraphics[width=0.85\textwidth]{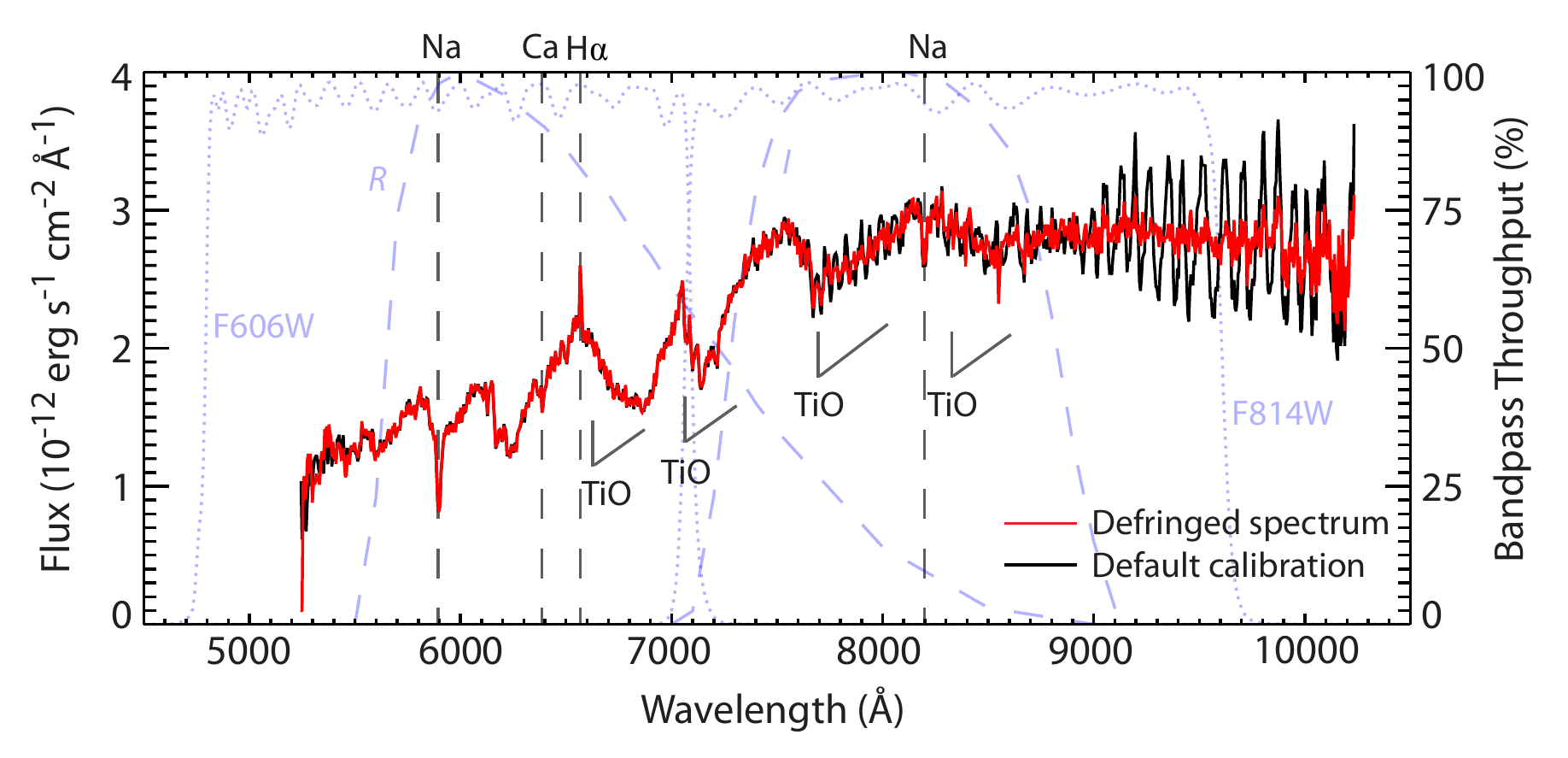}
\caption{AU Mic point source spectra with (red) and without (black) defringing. Major spectral features are labeled by the species that forms the line and marked by gray dashed lines and v shapes. The effects of reflections off the front and back surfaces of STIS's CCD can easily be seen in the default calibration (black) of our spectra at long wavelengths. For comparision, the througput of the F606W, F814W, \textit{R} and \textit{I} bands are displayed as the light blue dotted (\textit{HST} bands) and dashed curves (ground based bands) and are labeled as such (Bessell 1990; Ubeda et al. 2012). }\label{PointSource}
\end{figure*}
\end{center}

Once our spectra were defringed, we used the \texttt{STSDAS} \textit{wavecal} \texttt{IRAF} task to wavelength calibrate both the point source and fiducial spectra. We then flux calibrated the point source spectra with the \textit{x1d} task and the fiducial spectra with the \textit{x2d} task. This converts the units of each pixel from counts to erg s$^{ -1}$ cm$^{ -2}$ \AA$^{ -1}$ arcsec$^{ -2}$ while also correcting for geometric distortions so that the wavelength and distance axes increase linearly. Figure \ref{PointSource} displays the final reduced version of the AU Mic point source spectrum. For comparison, we also display the default calibration, which does not include defringing, for this spectrum. The defringing process significantly improves the visibility of intrinsic spectral features above 7,500 \AA, while removing the interference pattern imprinted onto the data by the STIS CCD. 

\subsection{Hot and Cold Pixel Correction}

We corrected our fiducial spectra for hot and cold pixels following a similar sigma-clipping procedure to that laid out by \cite{Aki}. As part of that prescription pixels inside a clip box, whose size is determined by the shape of the spectra, are used to calculate both the spectrum and background noise, which are then combined in quadrature. This ``local noise" is then used to determine if any individual pixel should be flagged for correction. Pixels that varied from the median flux for the whole clip box by more than three times the local noise were replaced with the median flux. In our case, the size of the clip boxes were 5 pixels in the spatial (y) direction and 30 pixels in the wavelength (x) direction. This size mitigated missing hot or cold pixels because it accounted for the large variations across the images in the spatial direction and the significant, but not as large, variations in the wavelength direction.

For a given clip box, the spectrum noise was calculated by computing the standard deviation of the values for all pixels in the box. Similarly, the background noise is the standard deviation of a five by five box of pixels surrounding the pixel with the largest (or smallest) value inside the clip box. We note that we do not follow the background noise calculation presented in \cite{Aki}, who used the standard deviation of the column of pixels containing the largest (or smallest) value. This is because many of our hot and cold pixels come in pairs (or more) directly next to each other, and the resulting background noise is therefore skewed toward not flagging pixels which require correction. For comparison, if we directly follow the procedure as laid out by \cite{Aki} we would only correct 100 pixels, which is significantly below both the thousands of pixels \citet{Aki} corrected in their data and below the number clearly present in our data. Alternatively, if we were only to use the spectrum noise to determine which pixels should be corrected, we would flag upwards of 3\% of the image for correction. Our five by five background noise box produced a compromise between these two extreme approaches; approximately 1.5\% of pixels were corrected in each individual fiducial dataset. 

\begin{center}
\begin{figure*}
\centering
\includegraphics[width=0.9\textwidth]{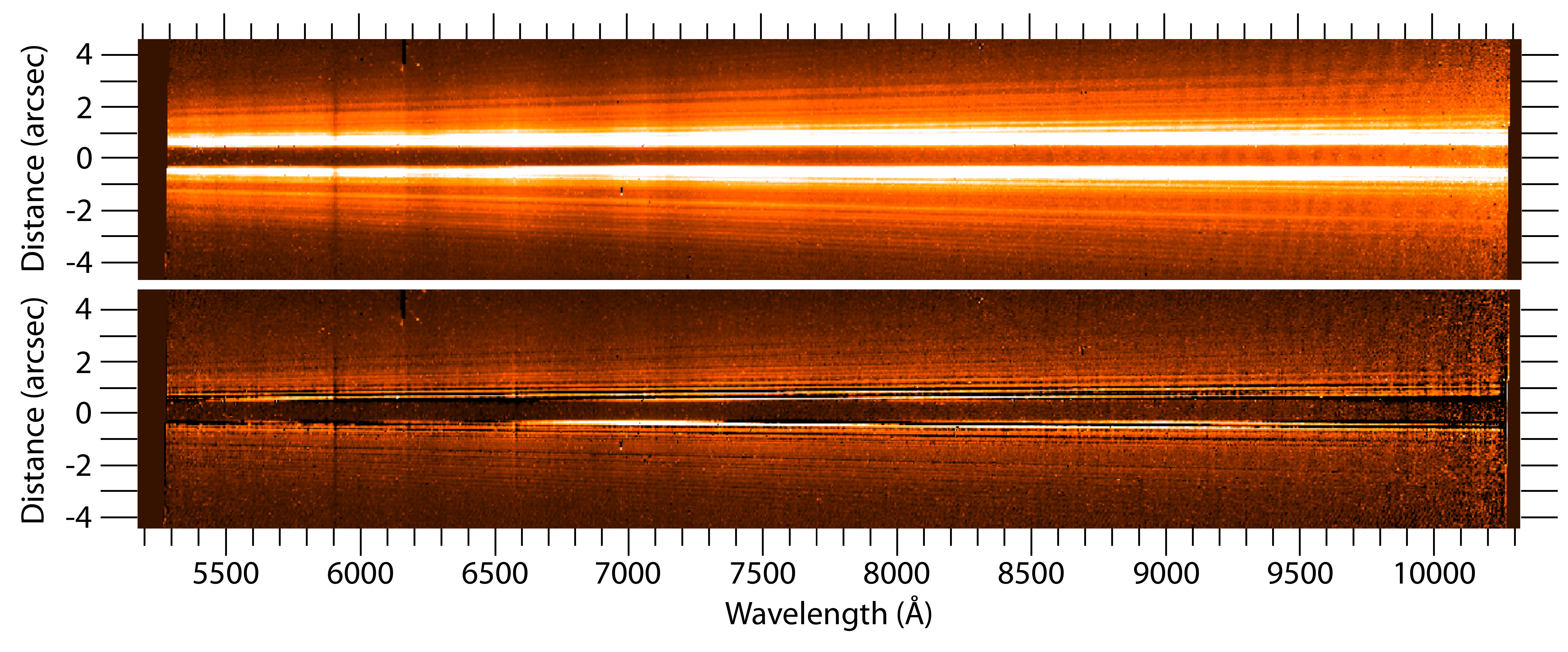}
\caption{Top: Sample individual (OBPZ05040) AU Mic fiducial spectrum after data reduction, and correction of hot and cold pixels. Some remaining hot and cold pixels are still visible, as well as the location of the fiducial bar (dark, horizontal feature), spectral features (dark, vertical features), and \textit{HST}'s Airy rings (bright lines angled away from the fiducial bar as wavelength increases). Bottom: Final sample individual (OBPZ05040) AU Mic fiducial spectra after PSF subtraction. Residuals from the PSF subtraction process can be seen close to the fiducial bar, as well as some residual fringing at long wavelengths. Both panels are displayed on the same scale, which was chosen to highlight the Airy rings and PSF residuals.} \label{Fiducial}
\end{figure*}
\end{center}

Figure \ref{Fiducial} (top) displays a reduced AU Mic fiducial spectrum (OBPZ05040) after applying our hot and cold pixel corrections. Significantly fewer hot and cold pixels remain after correction, and hot and cold pixels that contained spectral data were not corrected at a rate higher than pixels located where the disk was not detected. In fact, the pixels in the disk spectra were less likely to be corrected than background pixels. The fiducial bar is clearly seen as a dark, horizontal band which passes through the vertical center of the image. Spectral features, particularly the \ion{Na}{1} 5890 \AA\space line, are visible as the dark vertical bands in our final image, as well as \textit{HST}'s Airy rings, which appear as slanted lines that angle away from the fiducial bar.

\subsection{PSF Subtraction}

We observed the PSF star GJ 784 immediately after AU Mic using the same exposure sequence. These data were reduced using the same reduction procedures we applied to the AU Mic spectra. GJ 784 was chosen because of its close spectral match (M0V) to AU Mic (M1V) allows us to isolate light reflected in AU Mic's disk from direct light from the central star. We correct for the brightness difference and slight spectral mismatch between AU Mic and GJ 784 by using our point source spectra. Each column of the fiducial PSF image was scaled by the ratio of the flux of the AU Mic to PSF point source spectra at that same wavelength. Once the fiducial PSF image was scaled, we subtracted it from the AU Mic fiducial images. In addition to removing the Airy rings visible in Figure \ref{Fiducial}, which are due to the telescope's PSF, this also removed the flux in the spectrum due to AU Mic's central star, leaving a spectrum containing only light reflected in the disk.

Figure \ref{Fiducial} (bottom) displays the final PSF subtracted AU Mic fiducial image for observation OBPZ05040. Effects still present in the final dataset include some fringing at the longest wavelengths and PSF subtraction residuals near the fiducial bar. In general, these features are present within all three final fiducial images of AU Mic. The locations of the PSF residuals shift slightly in the wavelength direction and their intensity varies somewhat between images. We explored mitigating the presence of these features by scaling the brightness our final PSF fiducial spectrum by up to $\pm10\%$ of the brightness difference between AU Mic and GJ 784, but found this to be ineffective in reducing the residuals at best and to produce more residuals at worst. We also explored shifting the final PSF image relative the the AU Mic fiducial spectra by up to several pixels in both the wavelength and distance directions. While this did not necessarily produce additional residuals, it did not appear to decrease their presence in our final datasets. This process primarily shifted the location of the dark bands seen near the fiducial bar in Figure \ref{Fiducial} to larger and smaller wavelengths. Therefore, the reduced AU Mic fiducial spectra were PSF subtracted with no realignment of the images.

\section{Extraction of Disk Color Spectra}

\subsection{Individual Orbits: Do flares affect our dataset?}
In order to explore the extent to which flares from AU Mic's central star might affect our disk color determination, we extracted spectra from the fiducial observations obtained during each individual orbit separately. Because flares from M dwarfs primarily affect flux at higher energies, we expect that they would appear as an increase in the scattered flux on the blue end of our spectra. 



\begin{center}
\begin{figure*}
\centering
\includegraphics[width=\textwidth]{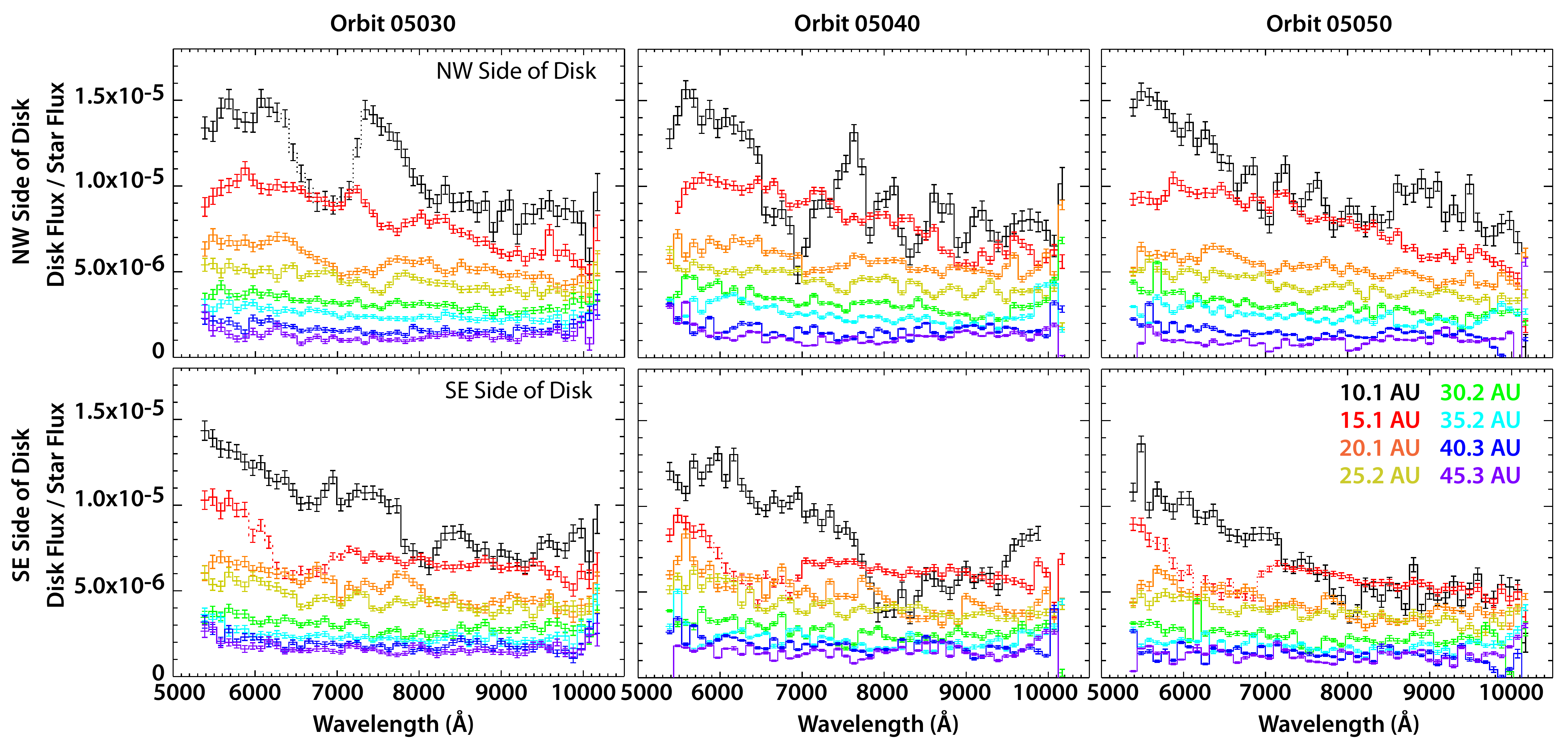}
\caption{Ratio of the combined AU Mic disk spectra to the stellar spectrum on the NW side (top panels; above the fiducial bar) and SE side (bottom panels; below the fiducial bar) of the system for our three orbits. Colors indicate spectra extracted at projected locations 10.1 AU (1\farcs 012; black), 15.1 AU (1\farcs 519; red), 20.1 (2\farcs 025; orange), 25.2 AU (2\farcs 531; yellow), 30.2 AU (3\farcs 037; green), 35.2 AU (3\farcs 544; light blue), 40.3 AU (4\farcs 050; dark blue), and 45.3 AU (4\farcs 556; purple) from the central star. The data have been binned by a factor of 20; the bin size is equivalent to approximately 98 \AA. The displayed error bars represent the combination of the statistical and 2\% systematic relative photometric uncertanties, which were added in quadrature (Riley et al. 2017). The dips between approximately 6000 \AA\space and 7000 \AA\space (dashed line) in the 15.1 AU spectrum on the SE side of the disk are due to PSF residuals. Regions of the 15.1 AU spectrum not affected by the PSF residuals are represented by solid lines. The 6500 \AA\space dip at 10.1 AU (dashed line) on the NW side of the disk appears in orbit 05030 to be associated with a stellar TiO line. The large scatter seen in the spectrum extracted from the NW side of the disk at 10.1 AU during orbit 05040 is due to PSF residuals near the fiducial bar. In general, the spectra from different orbits appear consistent with each other.}\label{IndOrbitSpectra}
\end{figure*}
\end{center}

We extracted spectra from every row of pixels for each of our three AU Mic fiducial exposures, which we corrected for slit loss using the \texttt{diff2pt} header keyword that was produced by the \textit{x2d} task. This is is a wavelength averaged throughput value such that the value is correct for the center of the bandpass, but too high and too low at the ends. We then converted from surface brightness to flux units by multiplying by the pixel scale in the dispersion direction (0.051 arcseconds) and the slit width (0.2 arcseconds). Finally, in order to produce our `color spectra', we divided our extracted disk spectra by our unocculted stellar spectrum to remove the edge effects introduced during our slit loss correction and produce a color relative to the system's central star. We calculated error-weighted mean spectra using a sliding bin in the distance direction that was 5 pixels (approximately 2.5 AU or 0\farcs 25) wide to reduce our uncertainties. Figure \ref{IndOrbitSpectra} shows our representative resulting color spectra, which we binned (20 pixels or 98 \AA) in the wavelength direction. Our displayed uncertainties represent the combination of the statistical and systematic errors, which were added in quadrature. When comparing our fiducial spectra against themselves, their relative uncertainty is up to 2\% of the total flux \citep{STIS}. Because our data were taken in back to back orbits, residual fringing and the absolute flux calibration dominate the systematic uncertainties of our dataset, while factors like the instrument stability and variability in the time dependent photometric calibration are insignificant. Therefore, the relative photometric uncertainties of our dataset are more likely to be closer to 1\%. However, out of an abundance of caution and desire to over, rather than under, estimate our uncertainties, the error bars displayed in Figure \ref{IndOrbitSpectra} were calculated using the 2\% systematic uncertainty value.

All of the color spectra are largely consistent with each other at the same projected separation independent of the side of the disk and orbit during which the data were taken. In general, the color spectra are blue at small distances from the central star and become grayer as projected separation increases. Several strong features are clearly visible in the color spectra that were extracted from regions close to the fiducial bar. In particular, there is a strong minimum near $\sim 6500-7000$ \AA\space (black dotted line) at a projected separation of 10.1 AU (1\farcs 012) on the NW side of the disk that appears in data from our first orbit (05030). This feature is consistent with a stellar titanium oxide band that remains present in our spectra and should not be considered when determining disk color. Additionally, there is a feature at $\approx 6500$ \AA\space (red dotted line) at a projected separation of 15.1 AU (1\farcs 519) on the SE side of the disk that persists through all three of our orbits of data. Examination of many disk spectra extracted at projected separations between 10 and 17 AU (1\farcs 012 - 1\farcs 1772) on both sides of the disk show that these minima behave like Airy rings; they move to longer wavelengths as the distance from the fiducial bar increases. We explored rescaling and shifting the alignment of our PSF, but found we could never completely remove these residuals. Therefore, we did not realign or rescale our PSF to produce our final color spectra. These features should not be considered significant in terms of determining disk color or the effects of flares on our color determination. However, because these PSF residuals do not affect the whole 15.1 AU (1\farcs 519) spectrum, the disk to stellar flux outside of this feature can still be used to determine the disk's color.

A comparison of how our disk spectra evolve with time reveals no evidence of flaring activity that might bias our color results. 

\subsection{Combined Color Spectra}

\begin{center}
\begin{figure}
\centering
\includegraphics[width=0.48\textwidth]{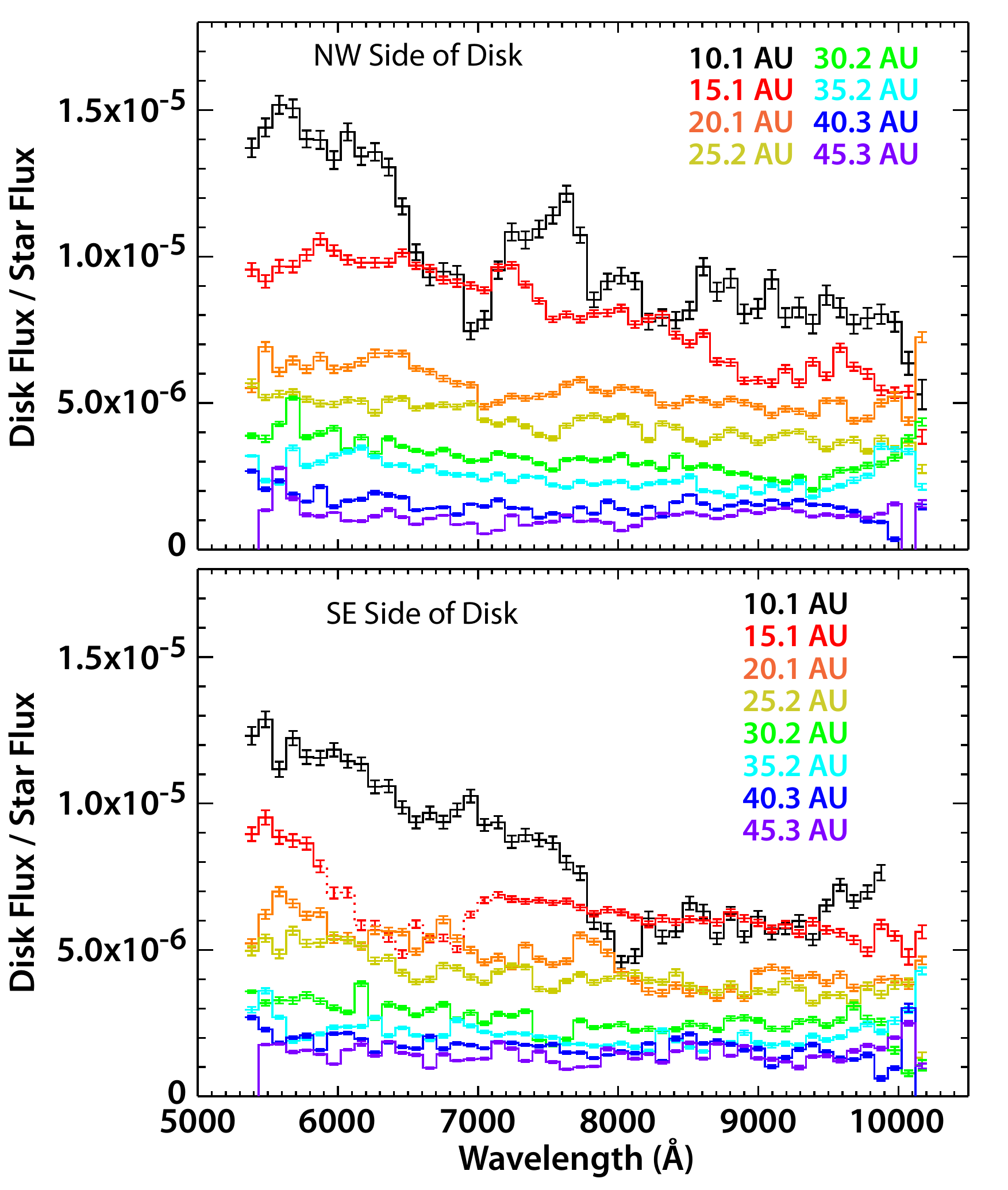}
\caption{Same as Figure 4, except for the final combined AU Mic spectra. In general, the spectra appear bluer close to the star and become increasingly gray as projected distance increases for both sides of the disk. The dip between 6000 \AA\space and 7000 \AA\space (dashed red line) in the 15.1 AU spectrum on the SE side of the disk is due to a PSF residual. Regions of the 15.1 AU spectrum unaffected by the PSF residual are indicated by solid lines.}\label{CombSpectra}
\end{figure}
\end{center}

We explored multiple ways to combine our three AU Mic fiducial datasets into one final fiducial spectrum. Ultimately, we found that combining these data after the reduction process produced the least amount of residual fringing. Therefore, each fiducial exposure underwent defringing, wavelength and flux calibration, hot and cold pixel correction, and PSF subtraction as described in Section 3 separately. We combined our individually extracted color spectra (see Section 4.1) into one final set of color spectra by calculating the error-weighted mean spectrum at each disk location. Figure \ref{CombSpectra} displays our final combined disk color spectra at representative projected separations on both the NW and SE sides of the disk. During the combination process we did not leave out data from regions in our individual spectra that we believe to be affected by TiO lines or PSF residuals. Instead, we combined the data from all the orbits and then reevaluated whether these effects appear in the final spectra. Our displayed uncertainties are the same as Figure \ref{IndOrbitSpectra} with the appropriate error propagation.

The final combined color spectra behave similarly to the color spectra extracted from individual orbits. The general trend in the data suggests the disk is bluer at small projected separations from the central star and, as distance increases, the disk becomes less blue on both sides of the disk. The PSF residual in the color spectrum extracted at 15.1 AU (1\farcs 519) on the SE side of the disk remains visible, but the effects from the TiO band at 10.1 AU (1\farcs 012) on the NW side of the disk is no longer evident.

We calculated the projected locations of the fast moving disk features B, C, and D, which were first detected by \cite{diskvar}, during our observations and extracted color spectra at those specific locations (15.10 AU, 26.67 AU, and 36.73 AU). We do not find any difference between the disk colors at these three specific projected disk locations and the disk colors from regions immediately surrounding these three locations. While our slit size was large enough to encompass some of these fast moving features, the apparent lack of color change associated with them may be partially due to our lack of vertical spatial resolution in the disk; spectra at the location of features B, C, and D includes light scattering off both the fast moving and other disk material. We did not attempt to determine the disk color for feature A (8.55 AU), because projected separations that close to the fiducial bar are significantly affected by PSF residuals, and feature E (50.32 AU), which is farther out in the disk than the maximum projected distance at which we have a spectral detection.

\section{Surface Brightness and Color Profiles}

\begin{center}
\begin{figure*}
\centering
\includegraphics[width=\textwidth]{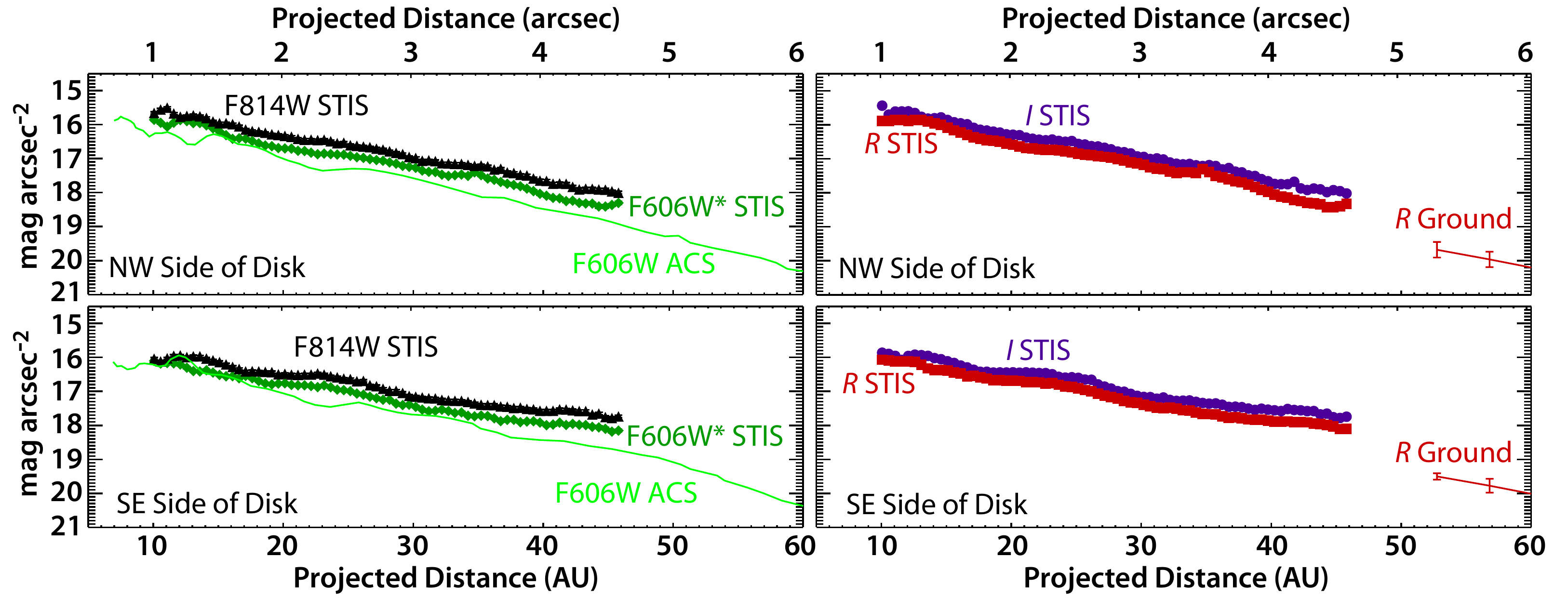}
\caption{Radial surface brightness profiles computed by convolving our coronagraphic spectra with the ACS F814W, ACS F606W, \textit{R}, and \textit{I} band filter response functions. Data from the NW side of the disk are displayed in the top panels, while data from the SW side of the disk are in the lower panels. The left panels display profiles for HST filters and the right panels display profiles for filters more typically used by ground based facilities. Profiles derived from our dataset are labeled by the filter response function used and 'STIS'. Our spectra do not cover the full wavelength range of the F606W filter. Therefore, we mark this filter with an asterisk. To facilitate comparison with datasets which currently exist in the literature, we over plotted the F606W ACS (green solid curves, left panels; \citealt{Krist}) and ground based \textit{R} band (red solid curves, right panels; \citealt{Kalas}) surface brightness profiles. We transcribed the F606W ACS data from Figure 4 in \cite{Krist}, which included no uncertainty measurements, and the \textit{R} band data from Figure 2 in \cite{Kalas}. In general, uncertainties are smaller than our point size and represent the combination of the statistical and 5\% systematic absolute photometry error, which were added in quadrature (Riley et al. 2017). The statistical uncertanty for each datapoint was assumed to be the throughput-weighted mean flux error.}\label{Comparison}
\end{figure*}
\end{center}

Prior to dividing our extracted disk spectra by the stellar spectrum, we convolved our spectra with synthetic filters (Figure \ref{PointSource}) to extract equivalent fluxes in the ACS F606W and F814W bands \citep{ACS}, as well as ground-based \textit{R} and \textit{I} bands \citep{Bessell}. This allows us to produce a pseudo-midplane surface brightness profile in each of these filters and facilitates comparisons between our dataset and other disk color datasets available in the literature, albeit with several caveats. First, we cannot calculate a true midplane surface brightness profile for our dataset because our data contain no spatial information in the vertical direction of the disk. Second, our disk spectra stop more than 500 \AA\space short of the blue end of the F606W filter. Third, there is no published color information for AU Mic's disk derived in part from \textit{I} band imagery. Additionally, while \cite{Krist} has published F814W imagery of the system, no detailed surface brightness profiles were presented. Finally, while \textit{R} band imagery of the system does exist \citep{Kalas}, it does not spatially overlap with the regions of the disk that we detect. All of our spectra are from disk locations which were under the occulting spot used by \cite{Kalas} in their discovery image. At regions outside of their occulting spot, the flux levels of our fiducial imagery are too low to detect the disk. Figure \ref{Comparison} displays our surface brightness profiles for all four filters for our final combined (error-weighted mean) dataset. In general, our surface brightness profiles show no distinct or significant features other than a general decrease in brightness as projected distance increases. The NW side of the disk is brighter than the SE side by, on average, 0.04 magnitudes in the \textit{R} band and 0.1 magnitudes in the \textit{I} band. Additionally, our profiles agree well with those from archival F606W and \textit{R} band datasets.

\begin{center}
\begin{figure*}
\centering
\includegraphics[width=\textwidth]{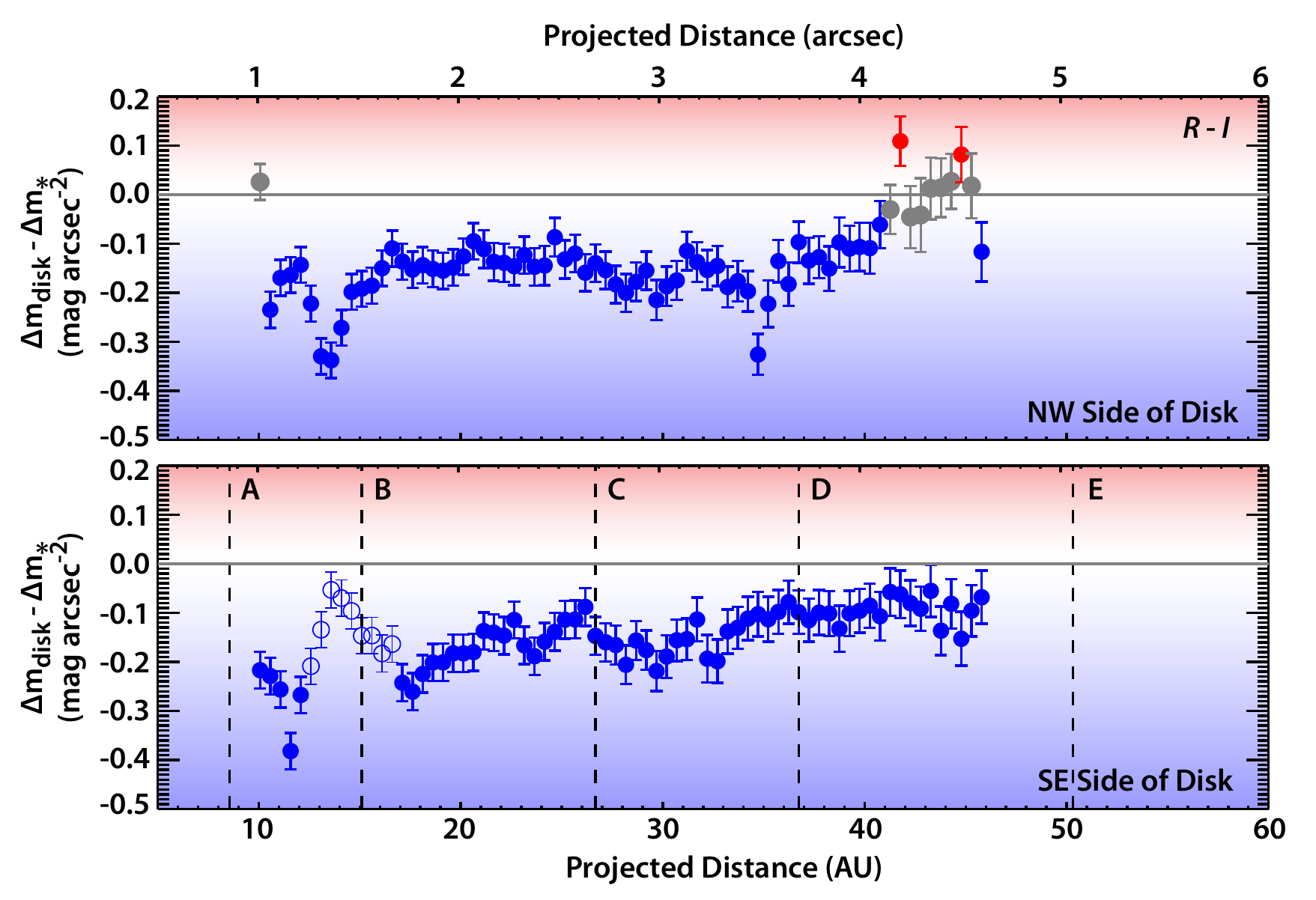}
\caption{\textit{R}-\textit{I} band disk color profile relative to AU Mic's central star for the NW (top) and SE (bottom) sides of the disk. The displayed error bars represent the combination of the 2\% systematic relative photometry and statistical uncertainties, which were added in quadrature. The disk is bluer for more negative color values, redder for more positive color values, and gray at a color of zero. We accentuate this with the background color gradient and the gray horizontal line at a color of 0.0 mag arcsec$^{ -2}$. Points are color coded to emphasize the disk color at those locations. Open circles represent data which are strongly affected by PSF residuals; trends in the data at these locations should be considered suspect and ignored. The locations of the fast moving features are marked by vertical dashed lines and labeled A through E using the same scheme as \cite{diskvar}. In general, the disk appears bluest in the inner disk and increasingly less blue as projected distance increases. Locations in the disk (30-60 AU) that were previously identified as the bluest are less blue in our dataset, suggesting disk color is variable. The data are suggestive of color changes associated with fast moving features C and D, but at a level that is not statistically significant.}\label{Color}
\end{figure*}
\end{center}

We computed \textit{R} minus \textit{I} band color information for the disk relative to AU Mic's central star using the surface brightness profiles displayed in Figure \ref{Comparison} and our unocculted AU Mic spectrum (Figure \ref{PointSource}), which we convolved with the same filter response functions. Figure \ref{Color} displays the resulting color profile for the NW and SE sides of the disk separately. The disk appears bluest from 10-17 AU on the NW and most of the SW side. The disk's color at larger projected separations is still modestly blue, albeit less so than at small projected separations. At projected separations larger than 40 AU, the disk color is consistent with being gray within 1$\sigma$ uncertainties at many locations on the NW sides of the disk.

On the SE side of the disk there is a sharp sudden graying of the disk around 14 AU (Figure \ref{Color} open symbols). This is due to the PSF residual that can be seen at the same projected distances in our color spectra (Figure \ref{CombSpectra}). Fast moving feature B \citep{diskvar} is located within the region of the our data affected by this PSF residual. Additionally, the locations of fast moving features A and E are outside of the regions of the disk that we detect. Sections of the disk immediately interior to features C and D appear to be less blue than regions immediately at and exterior to the fast moving features. It is difficult to discern the significance of this low level variability due to the size of our uncertainties. However, we note that our input spectra are binned using a moving box which is 5 rows of pixels wide in the original fiducial imagery. This has the effect of reducing the uncertainties on our disk color profile at the expense of reducing the amplitude of small changes in disk color such as those associated with the fast moving features.

On the NW side of the disk, there appears to be a peak in the blueness between 13 and 14 AU. This feature does not coincide with any obviously visible PSF residual in either our reduced fiducial imagery or extracted color spectra. Interestingly, \cite{Krist} reports a break in the ACS F606W surface brightness profile at a similar location in the disk ($\approx 15$ AU) which has not been observed at other wavelengths. Therefore, it is likely a real localized change in disk color. At projected distances immediately larger than this feature, the disk color appears to stay constant for approximately 15 AU before becoming increasingly less blue. The turnover to a less blue color occurs at approximately 30 to 35 AU on both sides of the disk, which is the same location where a second break in power law fits to surface brightness profiles have been reported (see e.g., F606W, J, H, and K' band imagery in \citealt{Liu,Metchev,Krist} and \citealt{Fitzgerald}).

We calculated the error weighted mean \textit{R}-\textit{I} band color for several regions of the disk to determine the significance of its change with projected distance. We also performed a linear fit of our color data to characterize its variations across those same regions. We report the slope of this fit, which is positive when the color of the disk is becoming less blue with projected separation, zero when the disk is uniformly blue, and negative when the disk is becoming more blue with increasing projected separation. Between 12 and 17 AU, the color of the NW side of the disk is $-0.214\pm 0.012$, while the slope of our linear fit is $0.024\pm 0.008$. This inner region of the disk is bluer than regions at immediately larger projected separations at the 3$\sigma$ level; between 17 and 35 AU the error-weighted mean color of the disk is $-0.154\pm 0.006$ and $-0.165\pm 0.007$ on the NW and SE sides respectively. Across these same regions the slope of our linear fit is $-0.004\pm 0.001$ and $0.004\pm 0.001$, respectively. This shows that the disk has not only become less blue in the 17-35 AU region when compared to the 12-17 AU region at a statistically significant level, but the variation in the color across these two regions is different at the 3$\sigma$ level as well. The blueness of the disk continues to decline at a statistically significant level. The error-weighted colors of the farthest projected regions (35-45 AU) of the disk that we detect are $-0.082\pm 0.011$ for the NW and $-0.096\pm 0.010$ for the SE sides of the disk. This is less blue than the regions of the disk at projected separations immediately interior to 35-45 AU at a $3\sigma$ level as well. On the NW side, the slope of our linear fit between 35 and 45 AU is $0.021\pm 0.004$, which is different at a 4$\sigma$ level from the 17-35 AU region on the same side of the disk. However, on the SE side, the disk becomes less blue at essentially the same rate for both the 17-35 AU and 35-45 AU regions; the slope of our fit at projected separations between 35 and 45 AU on the SE side of the disk is $0.001\pm 0.003$. This suggests that the trend of the disk becoming less blue with larger projected separations is real on both sides of the disk.

\begin{center}
\begin{figure*}
\centering
\includegraphics[width=\textwidth]{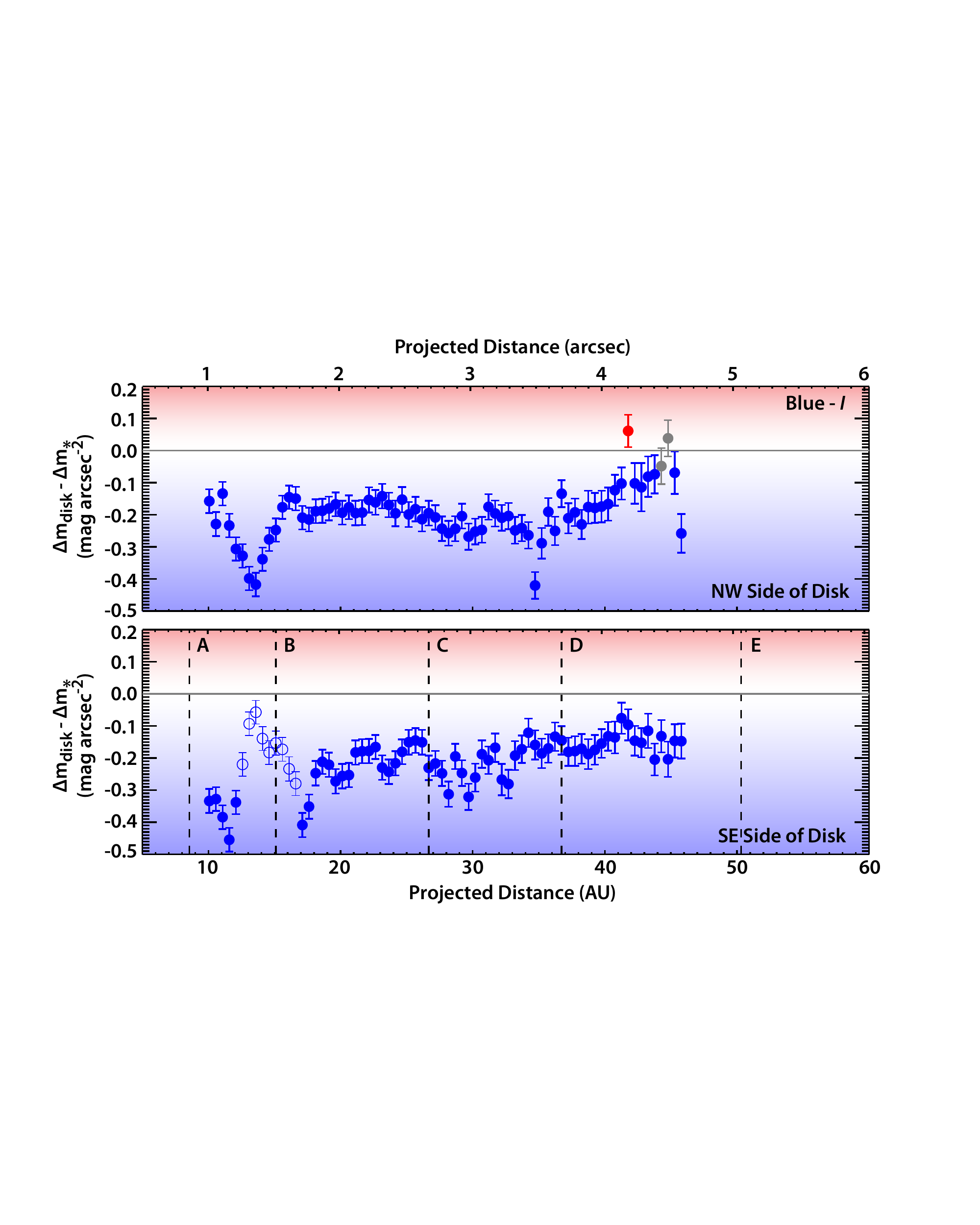}
\caption{Same as Figure \ref{Color} but for our Blue-\textit{I} band disk color profile relative to AU Mic's central star for the NW (top) and SE (bottom) sides of the disk. Our blue filter has 100\% throughput between 5500 and 7000\AA\space and 0\% throughput elsewhere. In general, the disk appears bluest in the inner disk and increasingly less blue as projected distance increases. Locations in the disk (30-60 AU) that were previously identified as the bluest are less blue than other regions of the disk in our dataset, suggesting disk color is variable.}\label{Color2}
\end{figure*}
\end{center}

The \textit{R} bandpass contains a large red tail that overlaps with the \textit{I} bandpass. This allows a significant amount of red light to be included in our \textit{R} band photometry and potentially impairs our color determination efforts. In order to more accurately trace the color, and thus the grain size distribution and chemistry, of the disk we created a `Blue' synthetic filter that has 100\% throughput between 5500 and 7000\AA\space and 0\% throughput everywhere else which we used to compute a second color profile for the system. Figure \ref{Color2} displays our resulting Blue-\textit{I} color profile. The same general trends exist in these profiles as the \textit{R}-\textit{I} profiles. The disk appears bluest in the inner disk and less blue in the outer disk. There appears to be a peak in the blueness of the NW side of the disk at 15 AU and a second break in the behavior of the color profile between 30 and 35 AU on both sides of the disk. Small changes in disk color at the locations of fast moving features C and D remain visible, but their significance is still difficult to determine due to the size of our uncertainties. At every separation the disk appears bluer in the Blue-\textit{I} color profile than the \textit{R}-\textit{I}. Beyond 40 AU the disk color no longer appears gray. This is due to the significant amount of red light `contamination' in the \textit{R} bandpass due to its long red tail.

\section{Discussion: Is the color of AU Mic's disk changing with time?}

The color of AU Mic's debris disk has been previously determined using a variety of high-contrast observations in optical to near-IR bandpasses. The optical (F435W-F814W; \citealt{Krist}), optical-near-IR (F606W-H; \citealt{Fitzgerald}, \citealt{Metchev}), and near-IR (J-H, K'-H \citealt{Fitzgerald}) colors of the disk become progressively bluer with projected separation at distances beyond $\sim$30 AU, except for the reported gray color within \textit{HST/ACS} F435W-F606W observations \citep{Krist}. At closer projected separations, the disk's color has been reported to be blue or gray and does not change with distance to the star \citep{Fitzgerald,Metchev}. The only possible exception to this is the qualitative report provided by \citet{Krist} that the F435W-F814W color becomes progressively bluer at smaller projected separations; however, these authors cautioned that strong PSF subtraction residuals rendered the trend suspect.  

By contrast, our optical coronagraphic spectroscopy appears to show slightly different color behavior than most previous broad-band color determinations of the disk. Both our color spectra and our color profiles (Figures \ref{CombSpectra}, \ref{Color}, and \ref{Color2}) show that AU Mic's disk is bluest in regions closest to its central star (projected separations between 12 and 17 AU; see Section 5). Thus, our data both quantify and confirm the trend qualitatively reported by \citet{Krist}. Moreover, these data seem to differ from mainly IR colors of the inner (10-30 AU) regions of the disk that find a more uniform blue or gray color with decreasing projected separation (see e.g. \citealt{Fitzgerald}). At more extended distances from the central star \text{(17-35 AU)}, the disk still appears blue, albeit less so than that observed closer to the star at a $3\sigma$ statistical significance. Moreover, the color of the disk at these extended distances (35-45 AU) does not become progressively more blue, as generally reported in previous literature, but rather is less blue than regions at smaller projected separations from AU Mic at a $3\sigma$ significance level.

Although our observations do not cover the full F606W bandpass and so do not allow us to make a complete comparison of our extracted colors with those quantified in the literature, our surface brightness data are generally consistent with archival imagery where comparisons can be made (Figure \ref{Comparison}). We have also carefully checked each of our \textit{HST} orbits to search for evidence of energetic flares, which as detailed in the Introduction are known to occur in AU Mic; however, we have found no clear evidence that the system experienced any significant flares during our observations. As far as we are aware, no other color dataset presented in the literature has been analyzed for such a signature outside of our own. We do note however that it would take a particularly strong flare to create such a signature.

Therefore, we consider the possibility that AU Mic's scattered light disk could exhibit variability in its observed optical colors. Variability in spatially resolved scattered light imagery of younger disks has been reported in several cases, and has generally been interpreted to arise from variable illumination of the outer regions of these disks induced by structure in the inner disk regions \citep{wis08,deb17}. AU Mic itself is also observed to have fast moving features in the SE side of its disk, whose origin is still unclear \citep{diskvar}. The color of a debris disk in scattered light can be influenced both by the grain size distribution present within the disk and by the detailed chemistry of these grains (e.g. \citealt{Debes}); hence, any purported color change must either be caused by a change in these grain properties or by a color change in illuminating source. Since we have already searched for and excluded the presence of large stellar flares during our observations and it seems unlikely that the chemistry of grains would change on the timescales involved, we suggest that the most likely cause of any time variable color change in AU Mic arises from a change in the grain size distribution.  

We speculate that the observed change in blue-optical colors of the disk that we observe at more extended projected separations (35-45 AU) could reflect a change in the grain size distribution that is scattering light. Specifically, the change in the disk's color from 30-45 AU from being ``increasingly bluer with distance'' \citep{Krist,Fitzgerald,Metchev} to a constant (SE side) and decreasing (NW side), small blue optical color could indicate a modest reduction in the relative number of sub-micron size grains being scattered at larger projected separations.  Analogously, the bluer optical colors that we are now clearly quantifying for the first time at projected separations of 12-17 AU could indicate an enhancement of sub-micron size grains inside of the purported ``birth ring'' of the system. However, the dust seen at small projected separations could also be the same dust belt observed by \cite{MacGregor} at 4\arcsec due to projection effects. Therefore, it is possible that the bluer optical colors observed at small projected separations may also be due to either viewing the same dust belt at differing scattering angles (e.g. a difference in color due to forward scattered light versus light scattered at a 90$\degree$ angle) or due to a scattering phase function with a slight wavelength dependence. While beyond the scope of this paper, detailed modeling of the scattered light from AU Mic's debris disk is needed to distinguish between the scenarios that may be responsible for the bluer color observed at small projected separations.

Identifying the origin of the purported change in the population of sub-micron size grains at more extended projected separations (30-45 AU) will likely require detail modeling of the dynamics of the AU Mic disk, which is beyond the scope of this paper. We do note that the dynamical state of the AU Mic disk was fundamentally different during the original epoch (2004) of \textit{HST} (F606W) observations, that have generally set the optical color baseline of the disk \citep{Krist}. Specifically, during the 2004 observations, only the fast, outward moving feature ``E'' identified by \citet{diskvar} had just begun to enter this region of the disk; its extrapolated location in the disk during 2004 was $\sim$34 AU, at a projected separation just inside of the reported increase in blue color. Features ``D'' and ``C'' were at extrapolated locations of $\sim$21 and $\sim$16 AU respectively during 2004. By contrast, feature ``E'' had moved beyond the extent of our detection capabilities ($\sim$50 AU) in our 2012 STIS coronagraphic spectroscopy data, as illustrated in Figures \ref{Color} and \ref{Color2}.  Moreover, a second fast, outward moving feature ``D'' ($\sim$37 AU) had moved within the 30-45 AU range by the epoch of our 2012 STIS observations, and feature ``C'' ($\sim$27 AU) was nearing this range. We speculate that there might be a causal correlation between these fast moving features propagating through the disk and the color changes that we are seeing in our scattered light data. Specifically, we speculate that as the fast moving features propagate outwards, they could preferentially elevate at least some small sub-micron size grains to higher latitudes above the disk midplane, and therefore reduce the previously noted increase in blue color with increasing projected separation. We note that the change in disk color revealed by our data seems to happen on both the NW and SE-side of the disk, whereas the fast, outward moving features have only been reported on the SE-side of the disk to-date. 

\cite{Chiang} recently suggested that the fast moving features are due to collisional chain reactions that form in an ``avalanche zone", where AU Mic's birth ring intersects a secondary debris ring that was formed within the last several ten thousand years. They further suggest that this zone is at very small projected separations ($\sim$3.5 AU) and produces sub-micron sized dust that is then blown outward by the central star's wind. Interestingly, this may explain the bluer color of the disk at small projected separations ($<$15AU) that we detected in our coronagraphic spectra, although perhaps only on the SE side of the disk where our data are strongly affected by a PSF residual. Coronal mass ejections might also be responsible for the clearing out of some small sub-micron sized grains from the disk (e.g. \citealt{Osten}) and recent work by \cite{Sezestre} shows that the stellar wind may play a key role in the velocity profile of AU Mic's fast moving features. However, it is difficult to explain the bluer color of the disk at small projected separations on both the NW and SE sides while also explaining the time variability of the disk's color at larger projected separations, even with the combination of these scenarios. Regardless of the explanation, it is clear that AU Mic's disk is undergoing poorly understood complex physical phenomena that is changing the localized distribution of sub-micron sized grains on years long timescales. We encourage future observations to re-examine the optical colors of AU Mic's disk and to confirm the two new trends revealed by our data. We note that the fast outward moving features A,B,C,D, and E \citep{diskvar} would be located at projected separations of 12.6, 20.3, 32.9, 46.1, and 60.6 AU in 2017. If the propagation of these features through the disk influences the localized distribution of small grains as the features pass by, we would expect to see additional color changes in these regions of the disk in new coronagraphic spectroscopy data.

\section{Concluding Remarks}

We find that the color of AU Mic's debris disk is bluest at small (12-17 AU) projected separations.  These results both confirm and quantify the speculative findings qualitatively noted by \citet{Krist}, and are different than IR observations that suggested a uniform blue or grey color as a function of projected separation between 10-30 AU \citep{Fitzgerald}. Unlike previous literature that reported the color of AU Mic's disk became increasingly more blue as a function of projected separation beyond $\sim$30 AU (e.g. \citet{Krist} state that ``The ratios of these [surface brightness] profiles indicate that the disk becomes increasingly blue at larger radii, at least for $r=30-60$ AU."), we find the disk's optical color between 35 and 45 AU to be uniformly blue on the SE side of the disk and decreasingly blue on the NW side. We note that this apparent change in disk color at larger projected separations coincides with several fast, outward moving ``features'' recently noted by \citet{diskvar} that are moving through this region of the disk. We speculate that these phenomenon might be related, and that the fast moving features could be changing the localized distribution of sub-micron sized grains as they pass by, thereby reducing the blue color of the disk in the process. We encourage follow-up optical spectroscopic observations of the AU Mic to both confirm this result, and search for further changes in the disk color caused by additional fast moving features propagating through the disk.

\acknowledgments
This paper is based on observations made with the NASA/ESA Hubble Space Telescope, obtained at the Space Telescope Science Institute, which is operated by the Association of Universities for Research in Astronomy, Inc., under NASA contract NAS 5-26555. These observations are associated with program GO-12512.

{\it Facilities:} \facility{HST (STIS)}.

\end{document}